\documentclass[preprint,prb]{revtex4}

\usepackage{graphicx, amsmath}

\begin{document}

\newcommand{\tr}{{\rm Tr}}
\newcommand{\grad}{{\boldsymbol \nabla}}

\title{On compressive radial tidal forces}

\author{Marco Masi}
\email{marco_masi2@tin.it}
\affiliation{Via Tiepolo 36, 35131 Padova, Italy}

\begin{abstract} \vspace{20mm}
Radial tidal forces can be compressive instead of disruptive, a possibility that is frequently
overlooked in high level physics courses. For example, radial tidal compression can emerge in
extended stellar systems containing a smaller stellar cluster. For particular conditions the tidal
field produced by this extended mass distribution can exert on the cluster it contains compressive
effects instead of the common disruptive forces. This interesting aspect of gravity can be derived
from standard relations given in many textbooks and introductory courses in astronomy and can
serve as an opportunity to look closer at some aspects of gravitational physics, stellar dynamics,
and differential geometry. The existence of compressive tides at the center of huge stellar
systems might suggest new evolutionary scenarios for the formation of stars and primordial
galactic formation processes.
\end{abstract}

\small Am. J. Phys. 75 (2), pp. 116-124, February 2007. \normalsize \vspace{20mm}


\maketitle

\section{Introduction}

Tides are the manifestation of a gradient of the gravitational force field
induced by a mass above an extended body or a system of
particles. In the solar system tidal perturbations act on compact bodies
such as planets, moons, and comets.\cite{Dermott} On larger scales than the
solar system, as in a galactic or cosmological context, we frequently deal
with tidal deformations (and eventually even disruptions) of a stellar
cluster by a galaxy or in galactic encounters. The study of tidal fields in
the most simple cases can be done by analytically approximating the
potential field of a nonspherical mass distribution by a spherical (that is,
a point mass) potential. This approximation sometimes leads to gross
estimates and is inappropriate for the analysis of tidal fields within mass
distributions.

Stars can be considered as particles with a long mean free path, and the
clusters they form as collisionless fluid dynamical systems. On galactic or
cosmological scales a huge mass distribution can interact gravitationally
with a smaller system that travels through it or is inside of it,
without any direct contact taking place between its constituents. For
example, think of a globular cluster passing through a disk or near the
center of a galaxy, or a dust cloud orbiting inside a star cluster.
The tidal field produced by the larger system, acting on the
smaller system it contains, needs to be computed inside the mass
distribution. Tidal forces can in these cases behave very differently than
in the cases encountered in the solar system.

Although numerical simulations are usually necessary to study an extended mass distribution, the
analytic study of some representative, not too approximate models is sometimes possible. The
effect of tides due to an extended mass distribution has been an active field of research in the
last two decades, especially since high speed numerical computers became available. Typical cases
of interest are galactic encounters,\cite{encounters} globular clusters under the influence of
the galactic mass distribution,\cite{clusters,Innanen} and Oort cloud perturbations by the
galactic field.\cite{Heisler} However, most of these investigations were on disruptive tidal
forces.

We are accustomed to think of gravitation as a
force field that
decreases rapidly with the distance from the field source. This dependence
is not necessarily true inside a mass distribution producing the field where
the force can increase with the distance from the center of mass. A trivial
example is a spherical homogeneous mass distribution with radius
$R$ and constant density $\rho_{0}$. To an outside observer at distance
$r>R$ the force field is $F(r)=-GM/r^{2}$, with $M=\rho_{0}\frac{4}{3}\pi
R^{3}$, and the tidal force
$\partial F(r)/\partial r=2GM/r^{3}>0$, which because it is positive is
responsible for the typical tidal bulges, that is, disruptive effects. If
we move inside the same distribution, we obtain
$F(r)=-GM(r)/r^{2}=-\frac{4}{3}\pi G \rho_{0} r$, but $\partial F/\partial
r=-\frac{4}{3}\pi G \rho_{0} <0$. The tidal field has become compressive
because inside the body the gravitational strength increases with the
distance from the center of the sphere.

A constant density fluid system in gravitational equilibrium is an
unrealistic case. The effect of the compressive nature of tidal fields has
been studied in the context of disk shocking of star
clusters.\cite{Ostriker} Fast encounters with the galactic disk or bulge
cause gravitational shocks where the entire cluster expands, but with a
short period of contraction immediately after the shock. Another attempt to
understand this effect was made by Valluri\cite{Valluri} who studied
compressive tidal heating effects. The effects of the mean tidal field of
a cluster of galaxies on the internal dynamics of a disk galaxy traveling
through it was discussed, and the disk experiences a compressive tidal
field within the core of the cluster. Das and Jog\cite{Das}
investigated the same effect for compressed
molecular cloud dynamics. They suggested that the compressive tidal field in
the center of flat core early type galaxies and ultraluminous galaxies
compresses molecular clouds producing the dense gas observed in the center
of these galaxies.

The nature of the radial compressive tide remains largely unexplored. It is
important not only for its physical effects, but because it is a fundamental
aspect of gravitation itself. In the following we formulate a tidal field
theory in which this effect arises naturally and investigate its relevance
in some real systems such as stellar globular clusters and galactic centers.

\section{The construction and meaning of the Newtonian tidal tensor}

The
Newtonian tidal tensor is a useful tool for the study of tidal fields. The
use of the tidal tensor for this purpose is not new, although not common. It
has been employed for example in the study of the angular momentum growth in
protogalaxies and more recently for investigating the origin of angular
momentum in galaxies through tidal torquing (see for example,
Refs.~\onlinecite{White} and
\onlinecite{Porciani}). Here we analyze the tidal field inside a mass
distribution and show how the latter naturally contains regions of radial
tidal forces sign reversal.

Because the Newtonian tidal tensor is frequently employed in a somewhat
obscure way in the literature, it is useful to recall some of its
properties and see how it naturally leads to the common tidal textbook
formalism.

Once a coordinate system and a frame of reference is defined (typically
with its origin at the mass distribution's center of mass) and given the
mass distribution's gravitational force field $\mathbf{F}(\mathbf{R})$ (with
the vector $\mathbf{R}$ vector at a particle
of the tidally perturbed system), the tidal forces arise as the
manifestation of the gradient $\grad \mathbf{F}(\mathbf{R})$. The
tidal force induced by a mass distribution on another body or another system
of particles is defined as the difference
between the gravitational force that the mass exerts at that point, and the
mean gravitational force $\left< \mathbf{F} \right>$ it exerts on the whole
body or system
\begin{equation}
\mathbf{F}_{\mathrm{t}} (\mathbf{R}) =
\mathbf{F}(\mathbf{R}) - \left< \mathbf{F} \right>.
\end{equation}
It can be shown that
the mean gravitational force acting on the tidally perturbed system is
equivalent to the force acting at its center of mass, $\mathbf{R}_{\rm cm}$.
Hence without any loss of generality, we can write the tidal fields as
\begin{equation}
\mathbf{F}_{\mathrm{t}} (\mathbf{R}) = \Delta \mathbf{F} =
\mathbf{F}(\mathbf{R}) -
\mathbf{F}(\mathbf{R}_{\rm cm}).
\end{equation}

Once the gravitational force field is expressed in terms of the potential
$\Phi(\mathbf{R})$ as
$\mathbf{F}(\mathbf{R}) = - \grad \Phi(\mathbf{R})$, $\mathbf{F}_{t}$ is
completely determined if we introduce the second rank symmetric Newtonian
tidal force tensor\cite{comm1}
\begin{equation}
\tau_{ij} = - \frac{\partial^{2}\Phi}{\partial x_{i}
\partial x_{j}}, \label{tau}
\end{equation}
with $i,j = (1, 2, 3)$ and
$x_{1} = x,\, x_{2} = y,\, x_{3} = z$; $\tau_{ij}$ is symmetric due to the
conservative character of gravitational fields. If we use Einstein's
summation convention, the components of the tidal force $F_{t}$ in its
differential form is
\begin{equation}
dF_{t_{x_{i}}} = \frac{\partial F_{i}}{\partial x_{j}}
dx^{j} = - \frac{\partial^{2} \Phi}{\partial x_{i} \partial x_{j}} dx^{j} =
\tau_{ij} dx^{j}. \label{dft}
\end{equation}

The tensor $\tau$ is the Jacobian of the gravitational force field $\mathbf{F}$ and the negative
Hessian matrix of the scalar potential function $\Phi$. These analytic properties of $\tau$ give
information about the maximum, minimum, and saddle points of the potential surface $\Phi(x, y,
z)$.\cite{notation1} The negative of the trace of $\tau$ gives Poisson's equation, which relates
the potential $\Phi$ to the mass density $\rho$
\begin{equation}
\nabla^{2} \Phi = \frac{\partial^{2} \Phi}{\partial x^{2}} +
\frac{\partial^{2} \Phi}{\partial y^{2}} + \frac{\partial^{2} \Phi}{\partial
z^{2}} = 4 \pi G \rho (x, y, z).
\end{equation}
Poisson's equation is nothing more than a trace invariant quantity of the
tidal tensor matrix.

It is possible to formulate the gravitational potential $\Phi$ through Poisson's equation, and
therefore the force field of heterogeneous spherical, spheroidal, and disk-like mass
distributions.\cite{Binney} This formulation is not straightforward, especially for irregular and
complex mass distributions. From linear algebra we know that the trace of a Hessian $\tau$ of a
local surface patch is an invariant quantity under coordinate transformations and that it equals
the sum of its eigenvalues, $\lambda_{i}$, that is, $\tr(\tau) = \sum_{i} \lambda_{i}$.
Differential geometry defines twice the value of the mean local curvature (the mean value of
curvature along the axes) as $H = \frac{1}{2} \tr(\tau)$, and the Gaussian curvature is defined by
the product of the eigenvalues: $K = \prod_{i} \lambda_{i}$, or by the determinant $\det(\tau)$.
Hence, there seems to be a relation between the Newtonian tidal tensor, the curvature of the
potential surface it represents, and mass density and tidal forces, which is reminiscent of the
connection between spacetime curvature and matter in general relativity. This relation is not a
coincidence.

The Newtonian tidal tensor $\tau_{ij}$ is the classical counterpart of the
more general fourth rank Riemann spacetime curvature tensor of general
relativity, which is why it is called Newtonian. Those
familiar with general relativity know that the Riemann tensor,
$R^{i}_{kjl}$, is also known as the
\textit{tidal force tensor}. It can be shown\cite{Misner} that in the weak
field limit the Riemann tensor is the classical Newtonian tidal force
tensor, that is,
${R^{i}}_{00j} = \tau_{ij}$. Curvature in general relativity is
not associated with Newton's inverse square force law, but rather
with the relative accelerations of neighboring particles (their geodesic
deviation), that is, with tidal forces. We might say that
spacetime curvature and tidal forces represent the same physical entity, or
to put it in Bondi's words,\cite{Bondi} ``Einstein's theory of
gravitation has its physical and logical roots \ldots in the existence of
Newtonian tidal forces.''

From this point on the Newtonian tidal tensor will be referred to as
the tidal tensor, implying, if not stated otherwise, the classical one.

To become better acquainted with the procedure of the next sections it is
be useful to recover the main concepts associated with the most typical
tidal perturbation -- that induced by the spherical potential of a point
mass\cite{comm2} $\Phi^{\rm pm}(r) = - G M/r$, where
$M$ is the mass of the point mass and $G$ the gravitational constant.
By applying Eq.~\eqref{tau} the tidal tensor in
Cartesian coordinates for an arbitrary frame of reference does not look very
elegant, because once we fix the coordinate origin, the tidal tensor is
generally non-diagonal for any test particle in a gravitational force field.
Because it is a real symmetric matrix, there must be a frame of
reference for which $\tau^{\rm pm}$ is diagonal.\cite{comm3} If we choose a
frame of reference where the $x$-axis passes through the particle's position,
corresponding to the radial spherical coordinate, that is, $(x = r,
y = 0, z = 0)$, then $\tau$ is diagonal and becomes
\begin{equation}
\tau^{\rm pm} = G
\begin{pmatrix}
\frac{2\,M}{{x}^{3}} & 0 & 0 \\
0 & - \frac{M}{{x}^{3}} & 0 \\
0 & 0 & - \frac{M}{{x}^{3}}
\end{pmatrix}
\equiv G
\begin{pmatrix}
\frac{2\,M}{{r}^{3}} & 0 & 0\\
0 & - \frac{M}{{r}^{3}} & 0 \\
0 & 0 & - \frac{M}{{r}^{3}}
\end{pmatrix}.
\end{equation}
From Eq.~\eqref{dft} we obtain
\begin{subequations}
\label{firstord}
\begin{align}
dF^{\rm pm}_{t_{x}} & = \tau_{xx}^{\rm pm} dx =
\frac{2\,GM}{{r}^{3}} dx \label{firstord1} \\
dF^{\rm pm}_{t_{y}} & = \tau_{yy}^{\rm pm} dy = - \frac{GM}{{r}^{3}} dy \label{firstord2} \\
dF^{\rm pm}_{t_{z}} & = \tau_{zz}^{\rm pm} dz = - \frac{GM}{{r}^{3}} dz, \label{firstord3}
\end{align}
\end{subequations}
which represents the cubic dependence of the radial force, and
the orthogonal longitudinal and latitudinal differential tidal forces,
respectively.

The radial tidal force represents a change in the absolute value of the
force vector, and the longitudinal and latitudinal vectors represent an
angular variation of the components of the central force:
\begin{subequations}
\begin{align}
\tau_{xx}^{\rm pm} & = \frac{2 G M}{r^{3}} = \frac{\partial
\mathbf{F}(r)}{\partial r} \\
\tau_{yy}^{\rm pm} & = \tau_{zz}^{\rm pm} = -
\frac{GM}{{r}^{3}} = \frac{\mathbf{F}(r)}{r}.
\end{align}
\end{subequations}
The sign has
an important meaning. The positive sign (in this case that of the radial
tide) represents a tensile force and the negative one represents a
compressive effect.\cite{comm4}

From Eq.~\eqref{firstord1} the radial tidal force induced by a point mass is to first order in $x$
\begin{equation}
\label{induce} F_{tx}^{\rm pm} = \frac{2GM}{{r_{\rm cm}}^{3}} \Delta x,
\end{equation}
where $\Delta x$ is the distance of the particle undergoing that tidal force from the center of
mass of the perturbed body, and $r_{\rm cm}$ is the distance of this center of mass from the tide
raising point mass of mass $M$. Equation~\eqref{induce} is a common textbook expression for tidal
forces. From Eq.~\eqref{induce} we recover the celebrated Roche's limit in the usual way
\begin{equation}
\Delta x_{\rm Roche} = \Big( \frac{M_{\rm cm}}{2 M} \Big)^{1/3} r_{\rm cm}.
\end{equation}
The quantity $\Delta x_{\rm Roche}$, also called the tidal radius, represents the distance at which the
disruptive radial tidal force on a system of free particles (a perfectly fluid body) of mass
$M_{\rm cm}$ exactly opposes the system's self-gravity, $F(r) = - G M_{\rm cm}/\Delta x^{2}$, the
force that holds it together. Inside Roche's limit this body will be torn apart by tidal forces.

Historically the description of tidal fields beginning with tensor-like objects was less
successful than its potential or force field counterpart because a tensor field needs more
entities to be visualized (for example, at each point imagine the vectors representing the columns
of a matrix), while the concept of a scalar or a vector field is intuitively easier to grasp. But
the price paid might be worse because textbooks sometimes present lengthy and complicated
calculations to recover Eq.~\eqref{firstord}.

\subsection{The tidal field of the spherical heterogeneous mass distribution}

As a good example of tidal fields inside matter distributions, we study
the potential of a heterogeneous spherical mass distribution. We can
show\cite{Binney} that the resulting spherical potential
is given by
\begin{equation}
\Phi^{\rm sph}(r) = -4 \pi G \Big[
\frac{1}{r}\!\int_{0}^{r} \rho(r') (r')^{\,2} dr' + \!\int_{r}^{R_{\max}}
\!\rho(r') r' dr' \Big], \label{sphericphi}
\end{equation}
where $\rho(r)$ is the radial mass density and
$R_{\max}$ is the maximum extension of the spherical mass distribution (if
we assume that no cutoff radius exists we can set $R_{\max} =
\infty$). Simple geometric considerations tell us that the radial dependence
of the enclosed mass in a sphere of radius $r$ and mass density distribution
$\rho(r)$ is
\begin{equation}
M(r) = 4 \pi\!\int_{0}^{r} \rho(r') (r')^{2} dr',
\label{mass}
\end{equation}
and therefore we write Eq.~\eqref{sphericphi} as
\begin{equation}
\Phi^{\rm sph}(r) = -
\frac{G \, M(r)}{r} -4 \pi G\!\int_{r}^{R_{\max}}\!\!\rho(r') r' dr'.
\end{equation}

The first integral term on the right-hand side of Eq.~\eqref{sphericphi} is the
gravitational potential, whose derivative is in accordance with Newton's second theorem, which
states that the gravitational force on a body placed outside a spherical shell of mass $M$ is
equivalent to the force produced by a point mass at the center of that shell.\cite{comm5} The
second integral accounts for the effect of the external mass distribution. We do not necessarily
need to know the radial mass dependence $M(r)$ to calculate $\tau$. The potential $\Phi$ or the
mass density function $\rho(r)$ alone can give us all the required information.

The application of Eq.~\eqref{tau} to Eq.~\eqref{sphericphi} is not so
straightforward because the former has been defined in Cartesian coordinates
and the latter in spherical coordinates. We expect that the calculations
become easier if we rewrite the tidal tensor in spherical coordinates for a
spherically symmetric potential. We can transform
the tidal tensor, which is the negative Hessian operator
on a scalar function, into the same negative Hessian matrix in spherical
coordinates. (This kind of transformation requires the use of tensor algebra,
covariant differentiation, and a somewhat complicated and long calculation
with differential geometric methods as shown in the appendix.) The tidal
tensor for a spherical potential is
\begin{equation}
\label{tausph}
\tau^{\rm sph} =
\begin{pmatrix}
- \frac{\partial^{2} \Phi^{\rm sph}(r)}{\partial r^{2}} & 0 & 0 \\
0 & - \frac{1}{r} \frac{\partial \Phi^{\rm sph}(r)}{\partial r} & 0 \\
0 & 0 & - \frac{1}{r} \frac{\partial \Phi^{\rm sph}(r)}{\partial r}
\end{pmatrix}
\end{equation}
In spherical coordinates the trace elements of $\tau^{\rm sph}$ highlight the distinction between
a tidal force induced by a radial change of the magnitude of the force field and that due to an
angular one
\begin{subequations}
\begin{align}
\tau^{\rm sph}_{rr} & = - \frac{\partial^{2} \Phi^{\rm sph}(r)}{\partial r^{2}} = \frac{\partial
\mathbf{F}(r)}{\partial r} \\
\tau^{\rm sph}_{\theta \theta} & = \tau^{\rm sph}_{\phi \phi} =
F_{\theta}(r) = F_{\phi}(r) = - \frac{1}{r}
\frac{\partial \Phi^{\rm sph}(r)}{\partial r} = \frac{\mathbf{F}(r)}{r}.
\end{align}
\end{subequations}
The first and second derivatives with respect to $r$ of
Eq.~\eqref{sphericphi} are
\begin{align}
\frac{\partial \Phi^{\rm sph}(r)}{\partial r} & = \frac{4 \pi
G}{r^{2}}\!\int_{0}^{r}
\rho(r') (r')^{2} dr' = \frac{G M(r)}{r^{2}} \\
\frac{\partial^{2} \Phi^{\rm sph}(r)}{\partial r^{2}} & = -
\frac{8 \pi G}{r^{3}} \!\int_{0}^{r} \rho(r') (r')^{2} \, dr' + 4 \pi G
\rho(r) = - \frac{2
G M(r)}{r^{3}} + 4 \pi G \rho(r).\label{d2phi}
\end{align}

Newton's second theorem for the force field of a spherical potential emerges
automatically
\begin{equation}
F = - \frac{\partial \Phi^{\rm sph}(r)}{\partial r} = - \frac{G M(r)}{r^{2}}. \label{nf}
\end{equation}
The direct application of Newton's second theorem is sufficient to determine the force field, and no
explicit evaluation of the potential derivatives is necessary. However, when dealing with tidal
forces in the presence of local matter where $\frac{\partial M(r)}{\partial r} \neq 0$, we cannot
avoid the explicit evaluation of both derivatives and we obtain
\begin{equation}
\tau^{\rm sph} =
\begin{pmatrix}
\frac{2 G M(r)}{r^{3}} - 4 \pi G \rho(r) & 0 & 0 \\
0 & - \frac{G M(r)}{r^{3}} & 0 \\
0 & 0 & - \frac{G M(r)}{r^{3}}
\end{pmatrix}.
\label{tausph1}
\end{equation}

Unlike gravitational force fields it is important to keep in mind that tidal
fields depend on the inside and the outside matter distributions.
If we define $\overline{\rho}(r) =
\frac{M(r)}{\frac{4}{3} \pi r^{3}}$ as the enclosed medium matter
density, that is, the homogenous density that would result from a mass
$M(r)$ enclosed in the sphere of radius $r$, we can also write
\begin{equation}
\label{tausph2}
\tau^{\rm sph} =4 \pi G
\begin{pmatrix}
\frac{2}{3} \overline{\rho}(r) - \rho(r) & 0 & 0 \\
0 & - \frac{1}{3} \overline{\rho}(r) & 0 \\
0 & 0 & - \frac{1}{3} \overline{\rho}(r)
\end{pmatrix}
\end{equation}
Thus we can also say that the strength of the radial tidal forces is
proportional to the deviation of the local density from the enclosed
medium matter density $\overline{\rho}(r)$.

As expected, the trace of $\tau^{\rm sph}$ is the scalar $-4 \pi G \rho(r)$,
according to Poisson's equation. However, the tidal tensor furnishes more
information than Poisson's equation alone, because it also gives the
individual contributions along the three coordinate axes. The components of
$\tau^{\rm sph}$ depend on the local mass density. The angular variation in
$d
\theta$ or $d \phi$ leads to a corresponding directional change of the tidal
force vector, which, in a spherical mass distribution, is a movement along
iso-density lines.

\section{The phenomenon of the tidal force reversal: the
spherical mass distribution}

\subsection{The tidal field for a general mass distribution model and for some special cases}
\label{sectionlabel}

The spherical heterogeneous mass distribution shows an
interesting possibility that cannot occur in the traditional case of a
point mass or for
tidal fields outside mass distributions. For special conditions the tidal
radial component
$\tau^{\rm sph}_{rr}$ in Eqs.~\eqref{tausph},
\eqref{tausph1}, or \eqref{tausph2}, can change its sign. Moving
radially inside the mass distribution, the tensile radial tidal force (the
force responsible for the usual tidal bulges) can become weaker, eventually
become zero, and even change its sign, that is, become a compressive radial
tidal force. For the spherical mass distribution this possibility occurs
when the local density nears, equals, or becomes greater than the enclosed
medium matter density. In the next sections we will show how this behavior
is a feature of tidal forces inside mass distributions for many geometries.

The gravitational force never changes sign, but its derivative can, and
that can make a difference for tidal forces. It all depends on how the local
matter density function behaves: in some physical conditions we
may well have an increase of gravitational forces moving away from the
central regions.

We now study this aspect more closely for the spherical
heterogeneous mass distribution. In particular, let us examine the
sign reversal of the tensor component
$\tau_{rr}^{\rm sph} = - \partial^{2} \Phi^{\rm sph} (r)/\partial
r^{2} = \partial F_(r)/\partial r$, and see what it
represents. From Eq.~\eqref{d2phi} we see that a transition from a tensile
to a compressive radial force occurs when
\begin{equation}
\tau_{rr}^{\rm sph} = - \frac{\partial^{2}
\Phi^{\rm sph}(r)}{\partial r^{2}} =
\frac{8 \pi G}{r^{3}}\!\int_{0}^{r}\! \rho(r') (r')^{2} \, dr' - 4 \pi G
\rho(r) = 0.
\label{condition}
\end{equation}

An immediate solution is the case of the singular mass density function, $\rho(r) \propto1/r$,
which leads to the exact equality of Eq.~\eqref{condition} for any radial distance $r$.
Equation~\eqref{condition} implies that a spherical mass distribution with a mass density function
proportional to the inverse of the distance from its center exactly cancels the radial tidal
forces everywhere. For any radial movement the increase (decrease) of the mass
enclosed in a sphere of radius $r$ is such that it determines an increase (decrease) of the force
field that exactly balances its decrease (increase) because of this radial change: the net
gravitational force field is constant inside the whole mass distribution. The potential $\Phi(r)
\propto \Phi_{0} + c r$, with $\Phi_{0} \equiv \Phi(0)$ and $c$ an arbitrary constant, the force
$\mathbf{F} = \partial \Phi(r)/\partial r \propto c = \mbox{constant}$, and there is no gradient
of the force field, $\partial^{2} \Phi/\partial r^{2} \equiv 0$, that is, no tidal forces
anywhere. In principle, we could imagine multiple star systems, Oort clouds, or entire star
clusters moving inside such a mass distribution without being affected by a tidal perturbation.

The mass density function $\rho(r) \propto 1/r$ is interesting as a
theoretical limit but is unrealistic or at least not very common in
astrophysical contexts.\cite{comm6} It is a theoretical limiting case
between a situation that gives rise to tensile or compressive tidal forces.

For a mass distribution function $\rho(r) \propto 1/r^{\gamma}$,
Eq.~\eqref{condition} becomes
\begin{equation}
\tau_{rr}^{\rm sph} = -
\frac{\partial^{2} \Phi^{\rm sph}(r)}{\partial r^{2}} \propto \frac{4 \pi
G}{r^{\gamma}}
\frac{\gamma -1}{3 - \gamma} \leq 0,
\end{equation}
which is satisfied if and only if $\gamma
\leq 1$. Inside a spherical mass distribution with $\gamma \leq 1$, the
radial tidal forces are always compressive. The usual tensile radial tides
with their induced tidal bulges to which we are accustomed no longer exist.
One example is the homogeneous mass distribution for
$\gamma = 0$,
$\rho(r) = \rho_{0} = \mbox{constant}$, for which we obtain
\begin{equation}
\tau_{rr}^{\rm sph} = - \frac{\partial^{2}
\Phi^{\rm sph}(r)}{\partial r^{2}} = - \frac{4 \pi}{3} G \rho_{0}.
\end{equation}
In this case it is not the force field that is constant (it
grows linearly with $r$), but the radial tidal component, which is
a compression force.

For $\gamma < 1$, $M(r) \propto r^{\alpha}$ with $\alpha > 2$, that is, the rate of increase of
the inner mass enclosed in a sphere of growing radius $r$ is faster than the inverse of the square
root law. If gravitational forces increase with $r$, the derivative has the opposite sign from
what we are accustomed to seeing for external or pointlike masses: it is this fact that leads to
the tidal reversal which is nonexistent for any field external to a mass distribution. We will see
that it is a common occurrence in the central regions of globular clusters, elliptic galaxies, and
galaxy bulges. Moreover, the notion of Roche's tidal radius is meaningless. Moving closer to this tidal reversal zone, the tidal
radius diverges toward infinity, and beyond this limit it does not exist. This analytic divergence
of the tidal radius should be taken into account when we consider stellar mass distributions,
which if neglected can lead to numerical errors.

\subsection{The tidal field for the isothermal mass distribution model}

So far we have analyzed some special cases that are unlikely in the real
world (though not completely unphysical). We will now consider cases with
$\gamma > 0$,\cite{comm7} where the central density does not go to infinity.

An interesting case of a real spherical heterogeneous mass distribution is that of globular
clusters.\cite{comm8} The surface brightness of these objects is well approximated by King's light
curve law.\cite{King} Therefore, if we assume that the visible bright matter represents the real
matter content (globular star clusters have a very low dark matter content if any), King's surface
brightness can be recovered by a three-dimensional isothermal mass distribution\cite{Binney}
\begin{equation}
\rho_{\rm iso}(r) = \frac{\rho_{0}}{1 + (
r/r_{c})^{2}} = \frac{\rho_{0}}{1 + \eta^{2}}, \label{rhoiso}
\end{equation}
where $\rho_{0}$ is the central density, $r_{c}$ a core radius (defined as the radius where the
surface brightness is 50\% of the central luminosity and is typically about the order of a parsec
in globular clusters and from $10^{2}$ to $10^{3}$\,pc for galactic bulges), and $\eta = r/r_{c}$
is a scale length.\cite{isothermal} The distribution $\rho_{\rm iso}$ is of order $1/r^{\gamma}$
with $\gamma \sim 0$ in the central regions and $\gamma \sim 2$ for the external ones. Thus, from
what we have seen in Sec.~\ref{sectionlabel}, we can expect a tidal reversal
phenomenon close to the central region.

We substitute Eq.~\eqref{rhoiso} into Eqs.~\eqref{sphericphi} and
\eqref{mass} and obtain the potential and the enclosed mass at the radius
$r$ for the isothermal mass distribution. If we scale all
formulas by the scale length $\eta = r/r_{c}$, we
obtain
\begin{equation}
\Phi_{\rm iso}(\eta) = - 4 \pi G r_{c}^{2} \rho_{0} \Big[\frac{\eta -
\arctan \eta}{\eta} -
\frac{1}{2} \log \Big( \frac{1 + \eta^{2}}{1 + \eta_{\max}^{2}} \Big)
\Big].
\end{equation}
We use Eqs.~\eqref{mass}, \eqref{tausph}, and \eqref{nf} for the
gravitational force field and the tidal tensor elements and obtain
\begin{align}
F_{\rm iso}(\eta) & = - 4 \pi G \rho_{0} r_{c} \frac{( \eta - \arctan
\eta)}{\eta^{2}} \label{Fiso} \\
\tau^{\rm iso}_{rr}(\eta) & = 4 \pi G \rho_{0} \Big[\frac{2(\eta - \arctan
\eta)}{\eta^{3}} -
\frac{1}{1 + \eta^{2}}\Big] \label{tauiso} \\
\tau^{\rm iso}_{\theta \theta}(\eta) & = \tau^{\rm iso}_{\phi \phi}(\eta) =
- 4
\pi G \rho_{0} \frac{ (
\eta - \arctan \eta )}{\eta^{3}}.
\label{tauisoang}
\end{align}

We see that $\lim_{\eta \rightarrow 0} \Phi_{\rm iso}(\eta) =
-2 \pi G r_{c}^{2} \rho_{0}
\log (1 + \eta_{\max}^{2})$ is finite if we assume there is a cutoff
radius $R_{\max} = r_{c}
\eta_{\max}$ for the dimension of the globular cluster. Similarly,
$M_{\rm iso}(\eta_{\max}) = 4 \pi
\rho_{0} r_{c}^{3} ( \eta_{\max} - \arctan \eta_{\max} )$ is
finite and represents the cluster's total mass. There is no singularity for
Eq.~\eqref{Fiso} because $\lim_{\eta \rightarrow 0} ( \eta -
\arctan \eta )/\eta^{2} = 0$, which is consistent with the
fact that gravitational forces must disappear at the center of a spherical
mass distribution. Note how for large $\eta$ it converges toward
an inverse force law instead of an inverse square one.

What do the gravitational and the tidal force fields look like in an isothermal mass distribution?
In Fig.~1 we show plots of Eqs.~\eqref{tauiso} (continuous line) and \eqref{tauisoang} (dashed
line), both normalized to $4 \pi G \rho_{0}$, versus the scale length $\eta$. The gravitational
force field increases with distance from the center until $\sim 1.5 \eta$. The radial tidal field
is correspondingly negative, which means that there are radial forces of compression in this
internal region. From Fig.~1 we see that there is a relatively intense negative peak for the
radial tidal field in the central part compared with the positive maxima at about $\sim 2.7 \eta$.
From this behavior it is clear how in the central regions of an isothermal sphere the radial tidal
compression is much greater than the tensile ones. From Eqs.~\eqref{tauiso} and \eqref{tauisoang}
we find that all the tidal tensor elements at the center of the cluster (that is, for $\eta
\rightarrow 0$) have a common value of $-4\pi G \rho_{0}/3$.

Compared with the positive maxima at about $\eta \sim 2.7$, which does not
exceed $\sim 0.03
\times 4 \pi G \rho_{0}$, it is clear how the compressive radial tidal
effects in a globular cluster's center might be more than ten times greater
than any radial tensile tidal force anywhere else. Moreover, it is
clear that there must be a small region inside such a mass distribution at
about $\sim 1.5$ scale lengths where the radial tidal effects are zero,
beyond which the negative tides take over abruptly.

\subsection{The tidal field for the Plummer sphere mass distribution model} \label{plummersection}

We briefly repeat the same calculations for the Plummer sphere model\cite{Binney} which is
sometimes used as a first approximation for the description of the internal structure of galactic
bulges. Plummer's model is interesting because it describes the simplest spherical potential that
does not diverge in the central regions as
\begin{equation}
\Phi_{\rm Pl}(r) = - \frac{G M_{T}}{\sqrt{r^{2}
+ r_{c}^{2}}},
\end{equation}
where $M_{T}$ is the total mass. To recover the mass density
function, $\rho_{\rm Pl}(r)$, we need to apply Poisson's equation directly
\begin{equation}
\nabla^{2} \Phi_{\rm Pl}(r) = \frac{1}{r^{2}} \frac{d}{dr} \Big( r^{2} \frac{d
\Phi_{\rm Pl}}{dr}
\Big) = 4 \pi G \rho_{\rm Pl}(r).
\end{equation}
Hence
\begin{equation}
\rho_{\rm Pl}(r) =
\frac{\rho_{0}}{(1 + \eta^{2})^{5/2}},
\end{equation}
with $\rho_{0} = 3 M_{T}/4 \pi
r_{c}^{3}$.
By using Eqs.~\eqref{mass}, \eqref{tausph}, and \eqref{nf} we see again that
the gravitational force field and the tidal tensor elements are
\begin{align}
F_{\rm Pl}(\eta) &= - \frac{4}{3} \pi G
\rho_{0} r_{c} \frac{\eta}{(1 + \eta^{2})^{3/2}} \\
\tau^{\rm Pl}_{rr}(\eta) &= \frac{4}{3} \pi G \rho_{0} \frac{2 \eta^{2} - 1}{(1 +
\eta^{2})^{5/2}} \\
\tau^{\rm Pl}_{\theta \theta}(\eta) &= \tau^{\rm Pl}_{\phi \phi}(\eta) = -
\frac{4}{3} \pi G \rho_{0} \frac{1}{(1 +
\eta^{2})^{3/2}}.
\end{align}

Plots of the tidal force fields are shown in Fig.~2. Its similarity with the isothermal mass
distribution is evident, but the gravitational force field is peaked more toward its center and
consequently there is a much more marked abrupt tidal reversal. Therefore it is expected that in
galactic centers the effect is less smooth than in globular clusters.

\section{Conclusion and possible directions of future research}

We have seen that expressing tidal forces in terms of a tensor formalism naturally leads to
understanding some features of a tidal field inside mass distributions. As an example of its
possible uses, we applied it to some astrophysical relevant cases. We saw how an inverse radial
tidal compressive effect emerges in the central regions of an isothermal model and a Plummer
model, which are first approximations for globular star clusters and galactic bulges mass
distributions.\cite{comm9} In these models the intensity of the central compressive tidal effects
depend only on the central density $\rho_{0}$, and the dimension of this central region is of the
order of the core radius. The most intense compressive tidal effects are expected for globular
clusters or galactic bulges with a relatively big core radius and a high central density.

These compressive tides may lead to dynamical effects (as might have occurred in the evolutionary
history of stellar clusters) in the form of gravitational tidal shocks,\cite{Ostriker} in tidal
heating,\cite{Valluri} in star forming clouds,\cite{Das} in binary or multiple star systems, in
the Oort clouds of comets,\cite{Masi} and is also expected in open star clusters when their
orbital path went through this internal region. For example, Oort clouds are dynamic systems that
can be very sensitive to the external galactic tides.\cite{Heisler} Galactic radial compressive
tides can reduce the Oort cloud volume considerably if it is inside the radial galactic tidal
reversal zone. But this compression would not lead to any gravitational collapse because the Oort
cloud comets can be considered as a non-self-interacting system of particles. Very different and
perhaps dramatic effects would cause a volume reduction of an interstellar cloud leading to a
gravitational collapse and to corresponding star forming process. There is now sufficient evidence
that suggests how some still to be determined, yet very intense star forming processes occur in
the center of galaxies up to the present. The tidal reversal phenomenon that occurs in the core
radius regions might be part of a possible explanation, despite the existence of massive black
holes whose disruptive tides would prevail only a few light years from its center. (Therefore, we
would expect a ring shaped star forming region.) Also, the transient and abrupt tidal compressions
we have seen at about the core radius might be considered as a source for a Jeans
instability.\cite{Jeans}

Isothermal models are also believed to approximate the dark matter halos of
galaxies. A galactic dark halo has a much lower central density (the order
of $\sim 0.1 M_{\odot}/pc^{3}$) than those found in globular clusters or
galaxy centers ($\sim 1000 M_{\odot}/pc^{3}$), and therefore any tidal
contribution
of a halo can be neglected. It is tempting to conjecture that
during the first cosmological phases, where the non-baryonic dark matter
density in the universe was much higher, if not dominant over bright
matter, a compressive instead of a disruptive tidal effect could have played
a role. This effect might have ignited a star formation process in some
regions of the universe much sooner than previously
expected, as has been observed at high redshifts. These lines of
research need further study.

\begin{acknowledgments}

I would like to thank Chris Hillman for his valuable suggestions on how to
calculate the tidal tensor in other coordinate systems.
\end{acknowledgments}

\appendix*

\section{The Hessian operator for a spherical coordinate system}

The nature of the Christoffel symbols necessary to calculate covariant
derivatives can be found in many textbooks. However these textbooks rarely
cover the case of the generalized fully covariant connection coefficients
that are necessary to obtain the Hessian in spherical coordinates. This
need is an opportunity to recapitulate covariant differentiation in a
broader context than usually given in introductory courses on differential
geometry.

If $\{x_{1}, x_{2}, x_{3}\}$ represent the Cartesian coordinate system and
$\{u_{1}(x_{1}, x_{2}, x_{3}), u_{2}(x_{1}, x_{2}, x_{3}), u_{3}(x_{1},
x_{2}, x_{3})\}$ represent a new one, then we might expect that by applying
the usual transformation law for a second rank tensor, the Hessian matrix of
a scalar function $\Phi$, $H_{\alpha \beta} = \partial^{2}
\Phi/\partial x_{\alpha} \partial x_{\beta}$, in these new set of
coordinates should become\cite{comm10}
\begin{equation}
H'_{\alpha \beta} = \frac{\partial^{2} \Phi}{\partial
u_{\alpha} \partial u_{\beta} } = H^{\gamma \delta} \frac{\partial
x_{\gamma}}{\partial u_{\alpha}} \frac{\partial x_{\delta}}{\partial
u_{\beta}}.
\end{equation}
Unfortunately, in general $\partial^{2}
\Phi/\partial u_{\alpha}
\partial u_{\beta}$ is not a tensor because this transformation does
not take into account the fact that the Hessian is the second derivative of
the gradient, which is not a vector but a covector. As it is known from
tensor analysis, to differentiate a covector requires covariant
differentiation. Thus, to recover the tidal tensor (the negative Hessian of
the potential) for other coordinate systems, we need to first calculate the
covariant derivative of the gradient (written also in the new coordinates).

The covariant derivative (usually labeled by the symbol ``;") of a covector
$T$ is
\begin{equation}
T_{\beta;\, \gamma} = T_{\beta, \gamma} - \Gamma_{\phantom{\alpha} \beta
\gamma}^{\alpha} T_{\alpha},\label{covdiff}
\end{equation}
where $T_{\beta, \gamma} $ is
the normal ``comma derivative'' on the component $\gamma$ of $T$, and
$\Gamma_{ \phantom{\alpha}
\beta \gamma}^{\alpha}$ are the connection coefficients of the second kind.
For Cartesian coordinates all the
$\Gamma_{\phantom{\alpha} \beta \gamma}^{\alpha}$ vanish and the covariant
derivative becomes the usual one. In this case the covector is the gradient
of $\Phi$ in the new coordinates $T_{\gamma} = \partial \Phi/\partial
u_{\gamma}$.

To calculate the connection $\Gamma_{\alpha\beta}^{\gamma}$ we need to distinguish between
coordinate bases (also called holonomic bases) and non-coordinate bases
(anholonomic bases). It is beyond the scope of this paper to give a
detailed and rigorous description of the differential geometric meaning of
these complications, and the interested reader is read the
relevant literature (for example, Ref.~\onlinecite{Misner}, Secs.~8.4
and 8.5). However, we can
clarify the difference briefly as follows.

Take the mathematician's viewpoint that tangent vectors and directional
derivatives are the same thing, $\textbf{u} = \partial_{\textbf{u}}$. Let
\textbf{u} and \textbf{v} be two vector fields and define their commutator
as
\begin{equation}
[\textbf{u},\textbf{v}] \equiv
[\partial_{\textbf{u}},
\partial_{\textbf{v}} ] =
\partial_{\textbf{u}} \partial_{\textbf{v}} - \partial_{\textbf{v}}
\partial_{\textbf{u}}.
\end{equation}
For any basis $\{\textbf{e}_{\alpha}\}$ we define the commutation
coefficients
${c_{\alpha \beta}}^{\gamma}$ by
\begin{equation}
[\textbf{e}_{\alpha}, \,
\textbf{e}_{\beta}] = {c_{\alpha \beta}}^{\gamma}
\textbf{e}_{\gamma}.\label{cc}
\end{equation}
By definition a basis is a ``coordinate basis'' or holonomic if
all the $c_{\alpha \beta}^{
\phantom{\alpha \beta} \gamma} = 0$ and a ``noncoordinate basis''
(anholonomic) if at least one
$c_{\alpha \beta}^{ \phantom{\alpha \beta} \gamma} \neq 0$.

For noncoordinate bases we calculate the connection
coefficients as
$\Gamma_{\phantom{\alpha} \beta \gamma}^{\alpha} = g^{\alpha \mu}
\Gamma_{\mu \beta \gamma}$, with
$\Gamma_{\mu \beta \gamma}$ the fully covariant connection coefficient of
the second kind given by
\begin{equation}
\Gamma_{\mu \beta \gamma} = \frac{1}{2} ( g_{\mu\beta,\gamma} +
g_{\mu\gamma,\beta} - g_{\beta\gamma,\mu} + c_{\mu\beta\gamma} +
c_{\mu\gamma\beta} - c_{\beta\gamma\mu}), \label{gamma}
\end{equation}
where $c_{\beta\gamma\alpha} =
g_{\alpha\mu} c_{\beta\gamma}^{\phantom{\beta\gamma}\mu}$.

For Cartesian coordinates $\textbf{e}_{x}\!\!=\!\!\frac{\partial}{\partial x}, \,\,
\textbf{e}_{y}\!\!=\!\!\frac{\partial}{\partial y}, \,\, \textbf{e}_{z}\!\!=\!\!
\frac{\partial}{\partial z}$, all commutation coefficients vanish, that is, it is a coordinate
basis, and Eq.~\eqref{gamma} simplifies to the usual Christoffel symbols.
For spherical coordinates
\begin{equation}
\textbf{e}_{\hat{r}}=\frac{\partial}{\partial r},
\quad \textbf{e}_{\hat{\theta}}=\frac{1}{r}\frac{\partial}{\partial
\theta},
\quad \textbf{e}_{\hat{\phi}}= \frac{1}{r \sin
\theta}\frac{\partial}{\partial \phi},\label{basis}
\end{equation}
Ref.~\onlinecite{Misner} obtain (p.\ 207)
\begin{align}
c_{\hat{r} \hat{\theta}}^{\phantom{\hat{r} \hat{\theta}
} \hat{\theta}} & = -
c_{\hat{\theta} \hat{r}}^{\phantom{\hat{\theta} \hat{r}
} \hat{\theta}} = -
\frac{1}{r}, \\
c_{\hat{r} \hat{\phi}}^{\phantom{\hat{r} \hat{\phi}
} \hat{\phi}} & = -
c_{\hat{\phi} \hat{r}}^{\phantom{\hat{\phi} \hat{r}
} \hat{\phi}} = - \frac{1}{r}, \\
c_{\hat{\theta} \hat{\phi}}^{\phantom{\hat{\theta}
\hat{\phi}} \hat{\phi}} & = -
c_{\hat{\phi} \hat{\theta}}^{\phantom{\hat{\phi} \hat{\theta}} \hat{\phi}}
= -
\frac{\cot\theta}{r}.
\end{align}
The connection coefficients for the spherical
noncoordinate basis given by Eq.~\eqref{gamma} are
\begin{subequations}
\label{gammaall}
\begin{align}
\Gamma_{\phantom{\alpha}
\hat{\theta} \hat{\theta}}^{\hat{r}} & = -
\Gamma_{\phantom{\alpha} \hat{r}
\hat{\theta}}^{\hat{\theta}} = - \frac{1}{r}, \label{gamma1} \\
\Gamma_{\phantom{\alpha} \hat{\phi} \hat{\phi}}^{\hat{r}} & = -
\Gamma_{\phantom{\alpha} \hat{r}
\hat{\phi}}^{\hat{\phi}} = - \frac{1}{r}, \label{gamma2} \\
\Gamma_{\phantom{\alpha} \hat{\phi}
\hat{\phi}}^{\hat{\theta}} & = - \Gamma_{\phantom{\alpha}
\hat{\theta}
\hat{\phi}}^{\ \hat{\phi}} = - \frac{\cot \theta}{r}, \label{gamma3}
\end{align}
\end{subequations}
and all other coefficients vanish.

To calculate the second derivatives of $\Phi$ (the Hessian), we must calculate the covariant
derivatives \eqref{covdiff}. If we write $H_{\beta\gamma}(\Phi) = \frac{\partial^{2}
\Phi}{\partial u_{\beta} \partial u_{\gamma}} = T_{\beta; \gamma}$ , with $T = \grad \Phi$ the
covariant vector to derive, then \begin{equation} H_{\beta \gamma} (\Phi) = (\grad \Phi)_{\beta,
\gamma} - \Gamma_{\phantom{\alpha} \beta \gamma}^{\alpha} (\grad \Phi)_{\alpha},
\end{equation} with $\Gamma_{\phantom{\alpha} \beta \gamma}^{\alpha}$ the coefficients \eqref{gammaall} in the
basis \eqref{basis}.
We obtain
\begin{subequations}
\begin{align}
H_{\ \hat{r} \hat{r}}(\Phi) & = ( \grad
\Phi)_{\hat{r}, \hat{r}} -
\Gamma_{\hat{r} \hat{r}}^{\alpha} (\grad
\Phi)_{\alpha} = \frac{\partial^{2} \Phi}{\partial r^{2}} \\
H_{\hat{r} \hat{\theta}}(\Phi) & =
H_{\hat{\theta} \hat{r}}(\Phi) =
( \grad \Phi)_{\hat{r}, \hat{\theta}} -
\Gamma_{\phantom{\alpha} \hat{r}
\hat{\theta}}^{\alpha} ( \grad \Phi)_{\alpha} =
\frac{1}{r} \frac{\partial^{2}
\Phi}{\partial \theta \partial r} - \frac{1}{r^{2}} \frac{\partial
\Phi}{\partial \theta}\\
H_{\hat{r} \hat{\phi}}(\Phi) & = H_{\hat{\phi} \hat{r}}(\Phi)
= (
\grad \Phi)_{\hat{r}, \hat{\phi}} - \Gamma_{\phantom{\alpha}
\hat{r}
\hat{\phi}}^{\alpha} ( \grad \Phi)_{\alpha} = \frac{1}{r
\sin \theta}
\frac{\partial^{2} \Phi}{\partial \phi \partial r} - \frac{1}{r^{2} \sin \theta}
\frac{\partial
\Phi}{\partial \phi} \\
H_{\hat{\theta} \hat{\theta}}(\Phi) & = ( \grad
\Phi)_{\hat{\theta}, \hat{\theta}} - \Gamma_{\phantom{\alpha}
\hat{\theta}
\hat{\theta}}^{\alpha} ( \grad \Phi)_{\alpha} =
\frac{1}{r^{2}}
\frac{\partial^{2} \Phi}{\partial \theta^{2}} + \frac{1}{r}
\frac{\partial \Phi}{\partial r} \\
H_{\hat{\theta} \hat{\phi}}(\Phi) & =
H_{\hat{\phi} \hat{\theta}}(\Phi) = ( \grad
\Phi)_{\hat{\theta}, \hat{\phi}} - \Gamma_{\phantom{\alpha}
\hat{\theta} \hat{\phi}}^{\alpha} ( \grad \Phi)_{\alpha} =
\frac{1}{r^{2}
\sin \theta} \frac{\partial^{2} \Phi}{\partial \phi \partial \theta} - \frac{\cos \theta}{r^{2}
\sin^{2} \theta} \frac{\partial \Phi}{\partial \phi} \\
H_{\hat{\phi} \hat{\phi}}(\Phi) & = ( \grad
\Phi)_{\hat{\phi}, \hat{\phi}} - \Gamma_{\phantom{\alpha}
\hat{\phi}
\hat{\phi}}^{\alpha} ( \grad \Phi)_{\alpha} =
\frac{1}{r^{2} \sin^{2} \theta}
\frac{\partial^{2} \Phi}{\partial \phi^{2}} + \frac{1}{r} \frac{\partial \Phi}{\partial r} +
\frac{\cot \theta}{r^{2}} \frac{\partial \Phi}{\partial \theta}
\end{align}
\end{subequations}

The Hessian matrix in spherical coordinates is written as
\begin{equation}
H(\Phi(r, \theta, \phi)) =
\begin{pmatrix}
\frac{\partial^{2} \Phi}{\partial r^{2}} & \frac{1}{r} \frac{\partial^{2}
\Phi}{\partial \theta \partial r} - \frac{1}{r^{2}}
\frac{\partial \Phi}{\partial \theta} & \frac{1}{r \sin \theta}
\frac{\partial^{2} \Phi}{\partial \phi \partial r} - \frac{1}{r^{2}
\sin \theta} \frac{\partial \Phi}{\partial \phi} \\
\frac{1}{r} \frac{\partial^{2} \Phi}{\partial r \partial \theta} -
\frac{1}{r^{2}}
\frac{\partial \Phi}{\partial \theta} & \frac{1}{r^{2}}
\frac{\partial^{2}\Phi}{\partial \theta^{2}} + \frac{1}{r} \frac{\partial
\Phi}{\partial r} & \frac{1}{r^{2} \sin \theta} \frac{\partial^{2}
\Phi}{\partial \phi \partial \theta} - \frac{\cos
\theta}{r^{2} \sin^{2} \theta} \frac{\partial \Phi}{\partial \phi} \\
\frac{1}{r \sin \theta} \frac{\partial^{2} \Phi}{\partial r \partial \phi} -
\frac{1}{r^{2} \sin
\theta} \frac{\partial \Phi}{\partial \phi} & \frac{1}{r^{2} \sin \theta}
\frac{\partial^{2} \Phi}{\partial \theta \partial \phi} - \frac{\cos
\theta}{r^{2} \sin^{2} \theta} \frac{\partial \Phi}{\partial \phi} &
\frac{1}{r^{2} \sin^{2} \theta} \frac{\partial^{2} \Phi}{\partial \phi^{2}}
+ \frac{1}{r}
\frac{\partial \Phi}{\partial r} + \frac{\cot \theta}{r^{2}} \frac{\partial
\Phi}{\partial
\theta}
\end{pmatrix}
\label{H}
\end{equation}
Note that the trace of $H$ equals the Laplacian of $\Phi$ in spherical
coordinates.

In general, Eq.~\eqref{H} is not a particularly useful or elegant expression for the Hessian.
However, if there is spherical symmetry, this matrix simplifies and the tidal tensor, which is the
negative Hessian of the gravitational potential $\Phi$, $\tau = - H$, applied to a spherical
potential $\Phi^{\rm sph}$ becomes the diagonal matrix \eqref{tausph}, as expected (all angular
derivatives $\frac{\partial}{\partial \theta}$ and $\frac{\partial}{\partial \phi}$ in
Eq.~\eqref{H} vanish).

If we are interested in describing tidal fields in elliptic or
disk-like galaxies, we can repeat the above calculation to obtain
the Hessian in elliptic and cylindrical coordinates respectively.

\newpage\section*{Figure Captions}

\begin{figure}[h]
\begin{center}
\includegraphics[width=125mm]{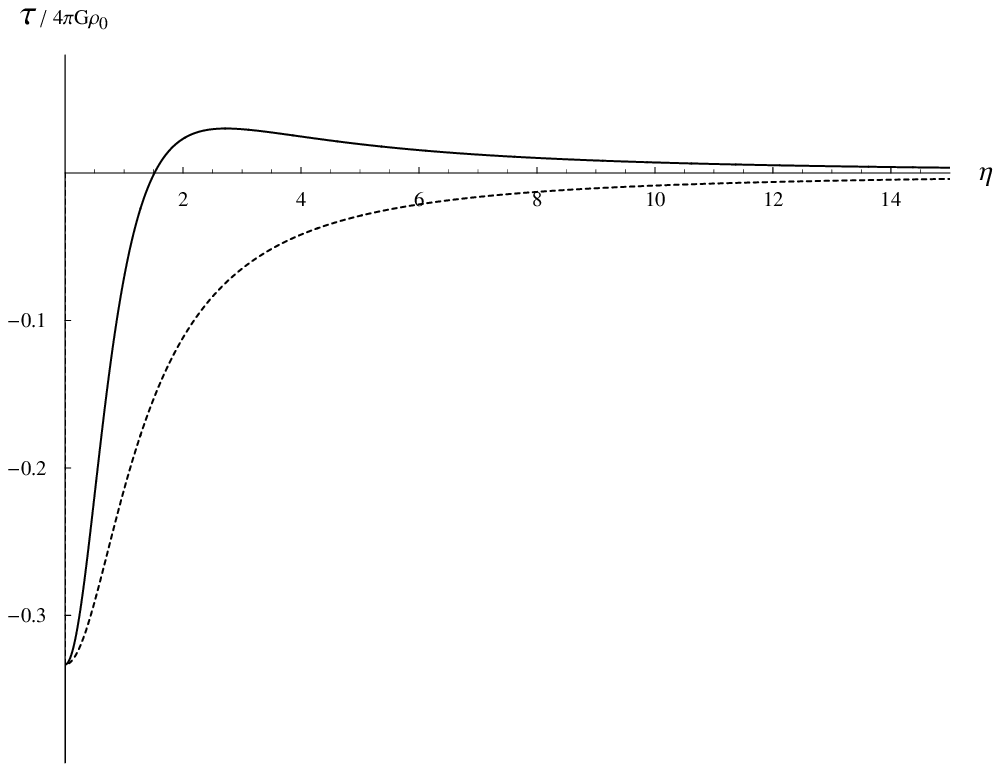}
\caption{The tidal force field of the isothermal spherical mass distribution
$\tau^{\rm iso}_{rr}(\eta)$ (continuous line) and $\tau^{\rm iso}_{\theta
\theta}(\eta) = \tau^{\rm iso}_{\phi
\phi}(\eta)$ (dashed line).}
\end{center}
\end{figure}

\begin{figure}[h]
\begin{center}
\includegraphics[width=125mm]{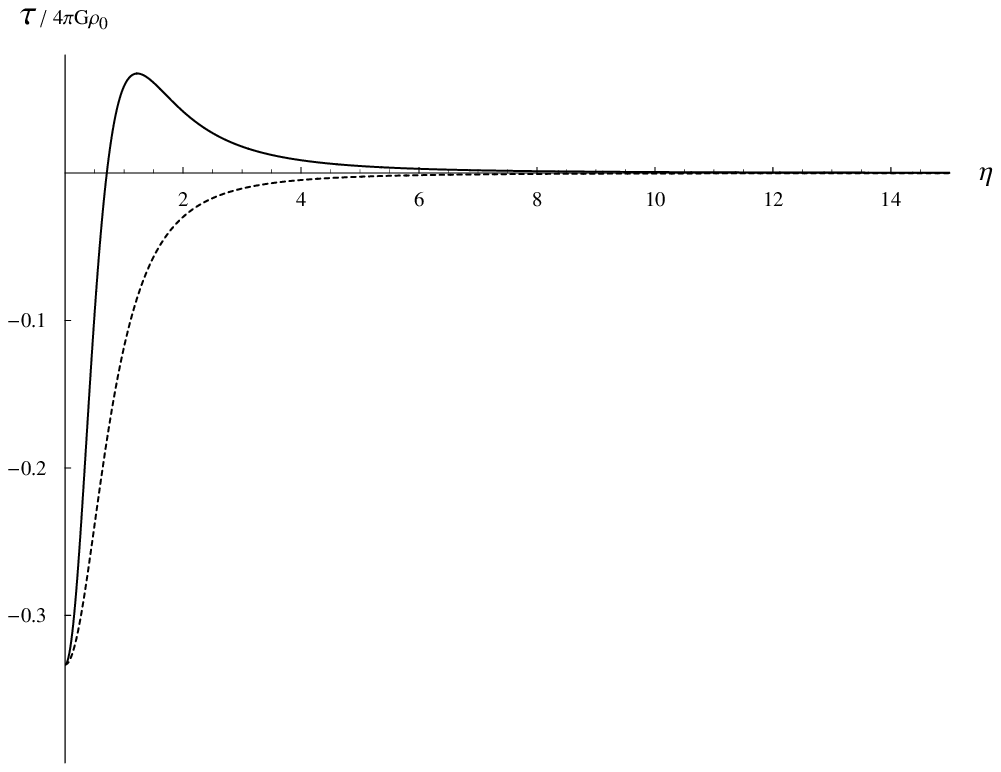}
\caption{The tidal force field of the Plummer spherical mass distribution
$\tau^{\rm Pl}_{rr}(\eta)$ (continuous line) and $\tau^{\rm Pl}_{\theta \theta}(\eta) =
\tau^{\rm Pl}_{\phi \phi}(\eta)$ (dashed line).}
\end{center}
\end{figure}


\begin{thebibliography}{}
\bibitem{comm1} In some of the literature the tidal tensor is identified
only with the pure shear deformation, that is, the traceless part of Eq.~\eqref{tau}, $\tau_{ij} -
\frac{\delta{ij}}{3} \tau_{ij}$, and Eq.~\eqref{tau} is called the ``deformation tensor.'' In
general, tidal deformations can be a combination of both shear and transverse stretch. We will
maintain the convention where both stretches are considered because, as we will see, it is more
consistent with the tidal tensor of general relativity.
\bibitem{notation1} When considering a generic, Cartesian, or radial coordinate system we will write
$\Phi(\mathbf{R})$, $\Phi(x,y,z)$, or $\Phi(r)$ accordingly.
\bibitem{comm2}We will use forces and potentials per unit mass.
\bibitem{comm3}From a purely algebraic point of view, we
should in principle calculate the eigenvalues of $\tau$ because every $n \times n$ real symmetric
matrix is orthogonally similar to a diagonal matrix whose diagonal elements are its eigenvalues.
Fortunately the diagonalization, at least in this case, is much simpler.
\bibitem{comm4}Some authors prefer to use the opposite sign. It is a
matter of convention. We follow this convention because it emerges directly
from the calculations without ambiguity.
\bibitem{comm5} Newton tried, without success, to show that this theorem also holds for homeoids
(ellipsoidal shells of matter), but the rigorous proof came about two centuries
later.\cite{Binney} Interesting applications and generalizations have been done recently also in
Ref.~\onlinecite{Secco}.
\bibitem{comm6}On the other side $\lim_{r \rightarrow 0} \rho(r) = \infty$
does not make this case unphysical if we think of the existence of a
singularity as a black hole at the center of the star cluster, and
especially notice how the mass does not diverge, because
$\lim_{r \rightarrow 0} M(r) = \lim_{r
\rightarrow 0} (2 \pi c r^{2}) = 0$.
\bibitem{comm7}The case $\gamma < 0$ on an interval in $r$ implies
an increasing density with $r$ which, even if not physically impossible (for example ring-shaped
matter distributions), is unlikely, or at least much less frequently in nature, and we will not
address it here.
\bibitem{comm8}A similar case has been studied by Valluri,\cite{Valluri} but
without transforming the tidal tensor to spherical coordinates. We will
compare this profile with Plummer's in Sec.~\ref{plummersection}.
\bibitem{isothermal} This density distribution corresponds to a stable arrangement of a self-gravitating system
that satisfies the equations of hydrostatic equilibrium of an ideal gas at a constant temperature.
\bibitem{Jeans} The Jeans instability occurs when the thermal pressure of a cloud of stellar gas,
which causes the cloud to expand, is no longer large enough to balance the gravitational pressure
that causes the cloud to collapse. In the scenarios we describe here we must consider not only the
self-gravitational pressure, but the tidal pressure must also be taken into account.
\bibitem{comm9} The models we have considered here are only approximations;
more realistic mass distribution might give more precise answers. However, the use of these
approximations does not undermine the fact that the tidal reversal is not an exception for a
specific model, but is a universal feature of the gravitational field.
\end{thebibliography}
\end{document}